\documentclass[aps,prc,twocolumn,superscriptaddress,showpacs,floatfix,nofootinbib]{revtex4-1}
\usepackage{amsmath,graphicx,color}
\newcommand{\trento}{T$\mathrel{\protect\raisebox{-2.1pt}{R}}$ENTo}

\begin{document}

\title{Relating centrality to impact parameter in nucleus-nucleus collisions}

\author{Sruthy Jyothi Das}
\affiliation{
Institut de physique th\'eorique, Universit\'e Paris Saclay, CNRS,
CEA, 91191 Gif-sur-Yvette, France} 
\affiliation{
Indian Institute of Science Education and Research (IISER)
Dr. Homi Bhabha Road, Pashan, Pune 411008, India}
\author{Giuliano Giacalone}
\affiliation{
Centre de physique th\'eorique, \'Ecole polytechnique, Universit\'e
Paris-Saclay, 91128 Palaiseau, France} 
\author{Pierre-Amaury Monard}
\affiliation{
Institut de physique th\'eorique, Universit\'e Paris Saclay, CNRS,
CEA, 91191 Gif-sur-Yvette, France}  
\author{Jean-Yves Ollitrault}
\affiliation{
Institut de physique th\'eorique, Universit\'e Paris Saclay, CNRS,
CEA, 91191 Gif-sur-Yvette, France} 
\date{\today}

\begin{abstract}
In ultrarelativistic heavy-ion experiments, one estimates the centrality of a collision by using a single observable, say $n$, typically given by the transverse energy or the number of tracks observed in a dedicated detector.
The correlation between $n$ and the impact parameter, $b$, of the collision is then inferred by fitting a specific model of the collision dynamics, such as the Glauber model, to experimental data. 
The goal of this paper is to assess precisely which information about $b$ can be extracted from data without any specific model of the collision. 
Under the sole assumption that the probability distribution of $n$ for a fixed $b$ is Gaussian, we show that the probability distribution of the impact parameter in a narrow centrality bin can be accurately reconstructed up to $5\%$ centrality.
We apply our methodology to data from the Relativistic Heavy Ion Collider and the Large Hadron Collider. 
We propose a simple measure of the precision of the centrality   determination, which can be used to compare different experiments.  
\end{abstract}

\maketitle

\section{Introduction}

The impact parameter, $b$, of an ultrarelativistic nucleus-nucleus collision is a crucial quantity. 
It determines the size and transverse shape of the quark-gluon matter formed in the collision. 
Central collisions, at small $b$, yield large and round interaction regions, while peripheral collisions, characterized by large values of impact parameter, yield smaller interaction regions with a pronounced elliptical anisotropy.
The centrality dependence of various observables provides, then, insight into their dependence on the global geometry. 
The energy loss of high-momentum particles~\cite{Adler:2002xw,Abelev:2006db} or jets~\cite{Aad:2010bu} is larger in central collisions, because it increases with the length of the path traversed by the particles inside the quark-gluon plasma.
By contrast, elliptic flow~\cite{Ackermann:2000tr,Aamodt:2010pa} originates from the elliptical shape of the nuclear overlap region, and is larger in peripheral events~\cite{Ollitrault:1992bk}. 

The impact parameter of a single collision, even though it is a perfectly well-defined quantity at ultrarelativistic energies (in the sense that the quantum uncertainty is negligible), is not directly measurable. 
In experiments, impact parameter is estimated by using a single observable, which we denote generically by $n$~\cite{Broniowski:2001ei}.
Depending on the experiment, $n$ is either the number of particles (multiplicity) in a given detector~\cite{Back:2000gw,Adler:2001yq,Adler:2004zn,Abelev:2013qoq} or the transverse energy deposited in a calorimeter~\cite{Chatrchyan:2011pb,ATLAS:2011ah}.
The idea is that collisions with a small impact parameter produce on average larger values of $n$. 
However, the relation between $n$ and $b$ is not one-to-one, and the variation of $n$ with $b$ is not known \textit{a priori}. 
This relation is usually inferred from a microscopic model of the  collision, such as {\scriptsize HIJING}~\cite{Wang:1991hta}, or a two-component Glauber model~\cite{Miller:2007ri} coupled with a simple model of particle production. 
The parameters of these models are tuned to reproduce the observed probability distribution of $n$.
While these models offer a convenient parametrization, they may not describe the actual dynamics of a collision. 
This is suggested by the fact that different sets of parameters must be used for different colliding systems, and by the observation that the two-component Glauber model is disfavored by analyses of U+U collisions~\cite{Adamczyk:2015obl,Moreland:2014oya}.

The goal of this article is to assess which information about the actual values of impact parameter can be extracted from the measured distributions of $n$, with as little theoretical bias as possible.
In particular, as we shall see in the following sections, we do not need to introduce the concept of ``participant nucleon'', which is a key ingredient of many microscopic models, but not a measurable quantity.



The term \textit{centrality} originally refers to a classification
according to impact parameter. 
In experiments nowadays, however, it refers to the classification of the collisions in terms of the parameter $n$. 
To avoid confusion, we call \textit{$b$-centrality} the centrality determined with respect to impact parameter. 
The corresponding definitions are recalled in Sec.~\ref{s:defcent}.
In Sec.~\ref{s:gaussian}, we show that a correspondence between $n$ and $b$ can be drawn under the sole assumption that fluctuations of $n$ for a given impact parameter are Gaussian.
This Gaussian is characterized by a mean $\bar n$ and a width $\sigma$, which both depend on impact parameter.
We test the validity of this assumption in Sec.~\ref{s:gaussian} by using model calculations.
We argue that data allow us to reconstruct unambiguously the full impact parameter dependence of the mean $\bar n$, and the value of the width $\sigma$ for central collisions, and we explain how this can be done in practice. 
In Sec.~\ref{s:trento}, we validate the proposed procedure by using model calculations, where the impact parameter is known. 
We show that the fluctuations of impact parameter at a fixed  centrality can be unambiguously reconstructed, and we apply this method to experimental data in Sec.~\ref{s:data}. 

\section{Two definitions of centrality}
\label{s:defcent}

Collisions can be classified according to their impact parameter, $b$. 
We define the centrality as the cumulative probability distribution of $b$: 
\begin{equation}
\label{defcb}
c_b\equiv \frac{1}{\sigma_{\rm inel}}\int_0^{b} P_{\rm inel}(b')2\pi b'db',
\end{equation}
where $\sigma_{\rm inel}$ is the inelastic nucleus-nucleus cross section and $P_{\rm inel}(b)$ is the probability that an inelastic collision occurs at impact parameter $b$. 
We name $c_b$ the \textit{$b$-centrality} of the collision, to distinguish it from the usual centrality defined in heavy-ion experiments, to be discussed below.  
The probability distribution of $c_b$ is flat by construction: $P(c_b)=1$ for $0<c_b<1$. 
Neither $b$ nor $c_b$ can be measured experimentally. 
They are known only in model calculations\footnote{The results in this paper use the variable $c_b$, but one can easily express them in terms of $b$ by using the change of variables  $c_b=\pi b^2/{\sigma}_{\rm inel}$. The value of $\sigma_{\rm inel}$ needs to be taken from either data or some collision model.}.

In experiments, collisions are instead classified according to a single observable, $n$. 
The STAR Collaboration~\cite{Adamczyk:2015obl} defines $n$ as the number of tracks of charged particles detected in the pseudorapidity window $-0.5<\eta<0.5$. 
The ALICE Collaboration~\cite{Abelev:2013qoq} uses the number of hits in two scintillators covering the windows $-3.7<\eta<-1.7$ and $2.8<\eta<5.1$.
The ATLAS and CMS Collaborations use the energy deposited in two forward calorimeters with symmetric acceptance windows: $4.9<|\eta|<3.2$ at ATLAS~\cite{ATLAS:2011ah}, and $3.0<|\eta|<5.2$ at CMS~\cite{Chatrchyan:2012vq}.

\begin{figure}[t!]
\begin{center}
\includegraphics[width=.95\linewidth]{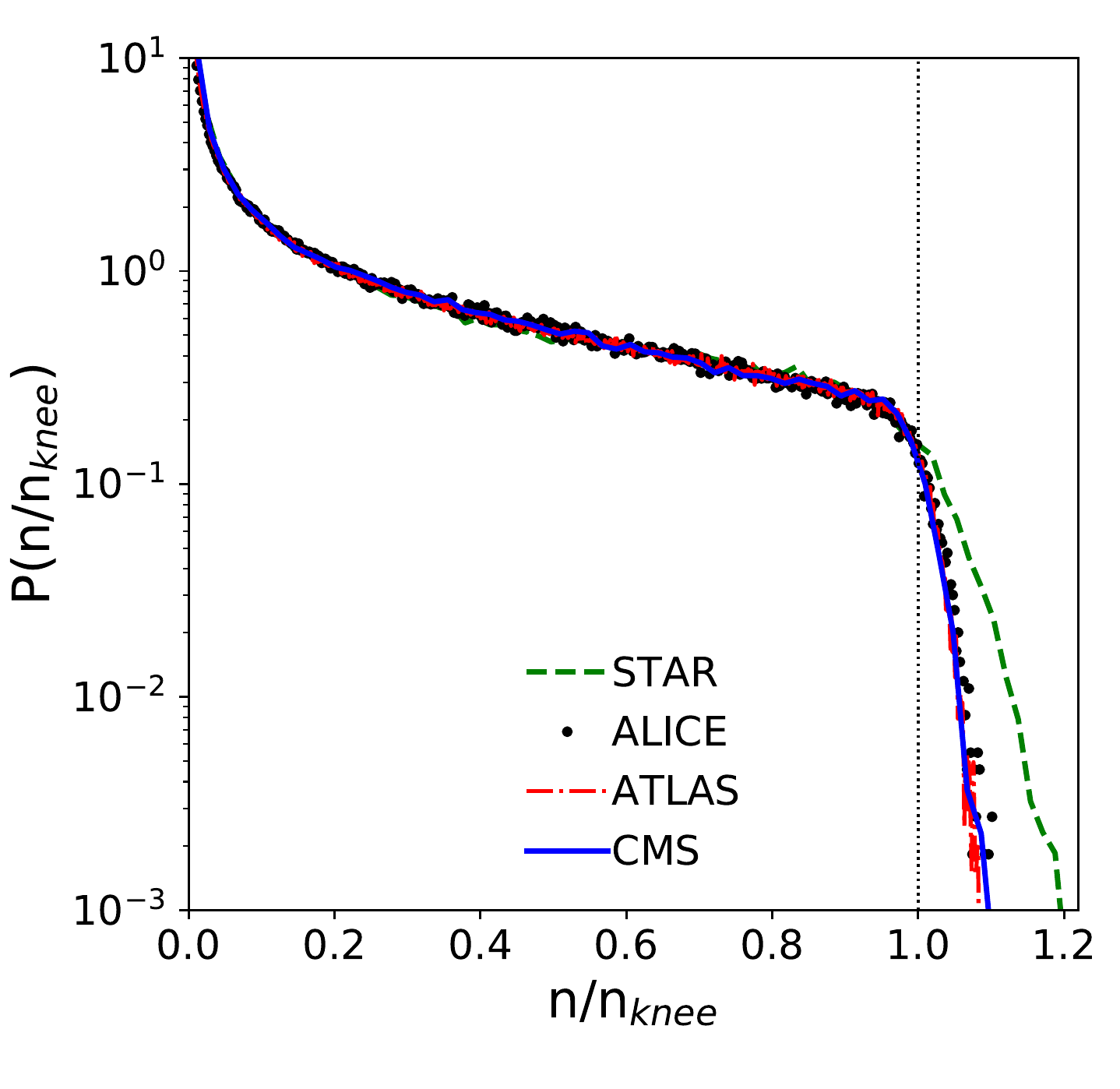} 
\end{center}
\caption{(Color online) Histograms of the probability distribution of $n$ measured by different experiments. Dashed line: STAR data on Au+Au collisions at $~\sqrt[]{s}=130~{\rm GeV}$ \cite{Adamczyk:2015obl}. The other curves show LHC data on Pb+Pb collisions at $~\sqrt[]{s}=2.76~{\rm TeV}$. Circles: ALICE data \cite{Abelev:2013qoq}. Dot-dashed line: ATLAS data \cite{ATLAS:2011ah}. Solid line: CMS data, extracted from Fig.~2 of Ref.~\cite{Chatrchyan:2012vq}. The horizontal axis of each histogram has been rescaled by the value of $n$ at the \textit{knee} (see text).}
\label{fig:allexp}
\end{figure} 
Figure~\ref{fig:allexp} displays the probability distribution of $n$, $P(n)$, measured by all these experiments.\footnote{Data on $P(n)$ collected by the PHENIX Collaboration can be found in Ref.~\cite{Adler:2004zn}.}
Since different detectors have different acceptance and efficiency, and $n$ can refer to a multiplicity or an energy, we rescale the value of $n$ by its value at the \textit{knee}, to be defined precisely in Sec.~\ref{s:gaussian}. 
Once rescaled, ALICE, ATLAS, and CMS data are almost identical.
STAR data differ in the tail, which is twice as  broad.\footnote{We use uncalibrated 130~GeV STAR data~\cite{Adler:2001yq} rather than calibrated 200~GeV data~\cite{Adamczyk:2015obl}. We have checked that, once rescaled, the two distributions are very similar. The advantage of 130~GeV data is that errors are provided, so that we are able to assess the quality of our fits, as we shall see in Sec.~\ref{s:data}.}
This difference can be ascribed to larger statistical fluctuations of multiplicity at RHIC. 
The pseudorapidity window used for the determination of centrality by the LHC collaborations is significantly larger than the window used by the STAR Collaboration, and the multiplicity per unit pseudorapidity is also smaller by a factor close to $2$ at RHIC than at the LHC~\cite{Aamodt:2010pb}. 
Therefore, the STAR detector observes fewer particles than the detectors at the LHC, which results in an increase of the relative statistical fluctuations.

The cumulative distribution of $n$ defines the experimental measure of centrality, which we denote by $c$. 
It is defined by 
\begin{equation}
\label{defc}
c\equiv \int_n^\infty P(n')dn'.
\end{equation} 
Note that the centrality classification is in ascending order for $b$ and in descending order for $n$, which explains the different integration limits in Eqs.~(\ref{defcb}) and (\ref{defc}). 
The probability distribution of $c$ is also flat by construction: $P(c)=1$ for $0<c<1$. 

We have thus defined two measures of the centrality: $c_b$ and $c$, depending on whether one sorts events according to $b$ or to $n$. 
If the relation between $n$ and $b$ is one-to-one, both measures coincide, $c=c_b$.  
In practice, one observes a range of values of $n$ at a given value of $b$. 
The joint distribution of $n$ and $b$ is usually inferred from a specific model of the collision~\cite{Wang:1991hta,Miller:2007ri}. 

\section{Relating centrality to $\boldsymbol{b}$-centrality}
\label{s:gaussian}

Here we simply assume that the probability of $n$ for fixed $b$ is Gaussian~\cite{Broniowski:2001ei}:  
\begin{equation}
\label{nfixedb}
P(n|c_b)=\frac{ \eta(c_b)}{\sigma(c_b)\sqrt{2\pi}}
\exp\left(-\frac{(n-\bar n(c_b))^2}{2\sigma(c_b)^2}\right),
\end{equation}
where both the mean, $\bar n$, and the width, $\sigma$, depend on the
impact parameter or, equivalently, on $c_b$, and
\begin{equation}
\eta(c_b) = 2 ~ \biggl[1+{\rm erf} \biggl( \frac{\bar n(c_b)}{\sigma(c_b)\sqrt{2}}\biggr)   \biggr]^{-1}
\end{equation}
normalizes the Gaussian to unity in the interval $0 < n < +\infty$.
Note that $\eta(c_b)$ is essentially equal to unity, except for very peripheral collisions.
We expect Eq.~(\ref{nfixedb}) to be a good approximation in a large system because of the
central limit theorem: $n$ is a multiplicity, or transverse energy,
which gets contributions from many collision processes which are  
located at different points in the transverse plane, and, therefore, causally disconnected and independent. 

It is useful to check this Gaussian approximation on a model.
The final-state observables used to define the collision centrality (e.g., multiplicity or transverse energy) are, in the hydrodynamic framework, proportional to the initial entropy of the system.
The initial entropy, then, corresponds to the experimental $n$, and is provided by models of initial conditions, such as the \trento{} model of initial conditions, which we use to simulate Pb+Pb collisions at $~\sqrt[]{s}=2.76~{\rm TeV}$.
The \trento{} model is a parametric model in which the entropy deposition is regulated by two parameters: $p$, which specifies the dependence of entropy deposition on the thickness functions, $T_A$ and $T_B$, of the incoming nuclei, and a parameter $k$ which governs the width event-by-event fluctuations of the entropy produced by each participant nucleon.
For a given choice of $p$, the value of $k$ can be tuned to match the distribution of the multiplicity of a given collision system~\cite{Moreland:2014oya}.
We choose $p=0$ (corresponding to a total entropy proportional to $\sqrt{T_AT_B}$ in each event), and $k=1.6$. 
This setup allows us to capture accurately a wide range of observables in Pb+Pb collisions at the LHC~\cite{Bernhard:2016tnd,Giacalone:2017uqx}.
We stress that our subsequent analysis of the experimental results does not rely on whether or not this setup of \trento{} provides a good description of data.

\begin{figure}[t!]
\begin{center}
\includegraphics[width=\linewidth]{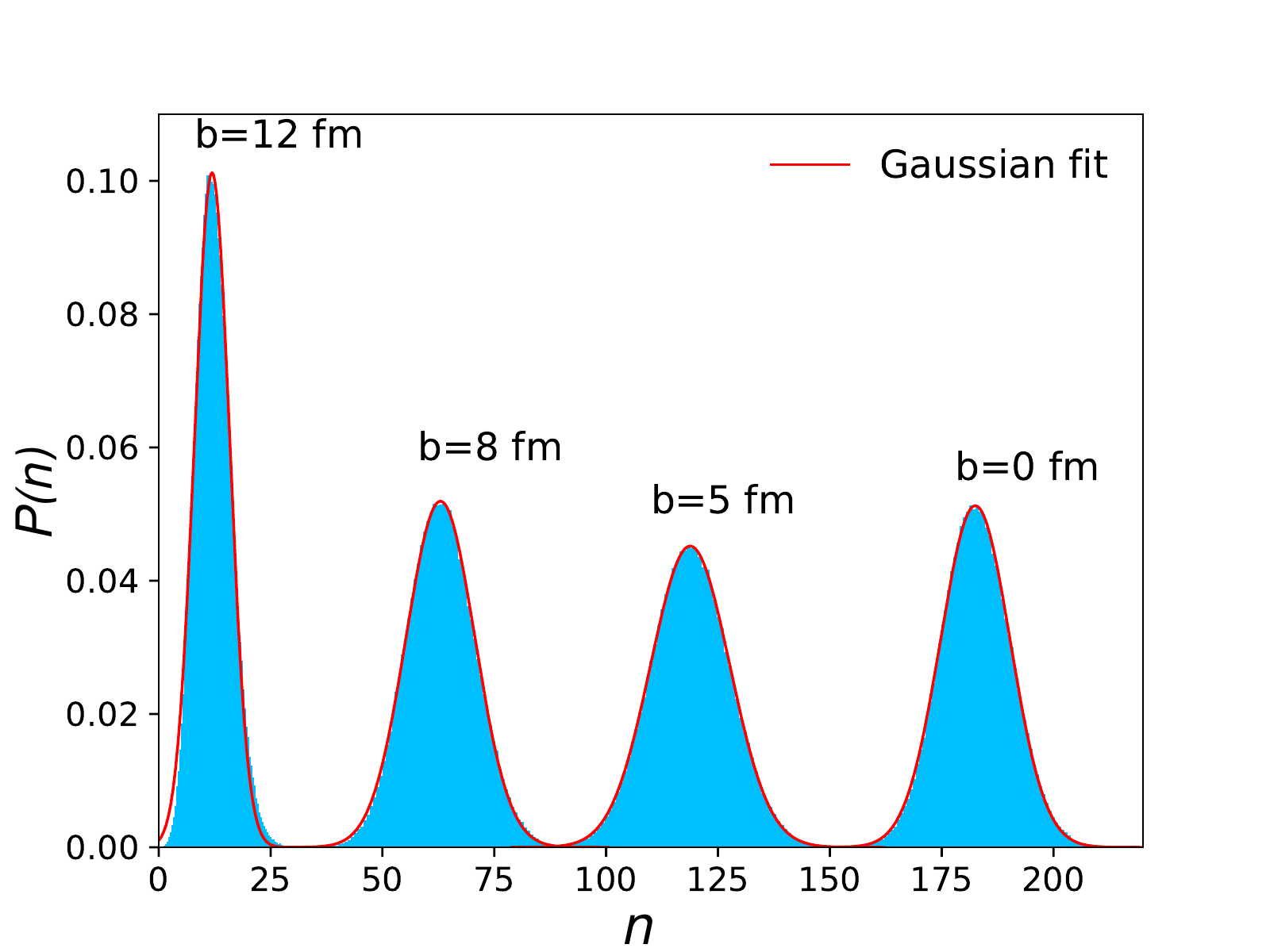} 
\end{center}
\caption{(Color online) Shaded areas correspond to histograms of the probability distribution of $n$ for fixed impact parameter in Pb+Pb collisions at $~\sqrt[]{s}=2.76~{\rm TeV}$ in the \trento{} model~\cite{Moreland:2014oya}. The values $b=0$ , 5, 8, 12~fm were used, which correspond to $c_b=0$, 10\%, 26\%, 58\%, respectively. We generated $5\times 10^5$ events for each value of $b$. Solid lines are Gaussian fits. The quantity $n$ is in arbitrary units.
}
\label{fig:gaussian}
\end{figure}

Figure~\ref{fig:gaussian} displays the distribution of total entropy in this setup of \trento{} for different fixed values of impact parameter.
This plot shows that the Gaussian approximation is valid in the model up to $\sim60\%$ $b$-centrality.
Around $b=12$~fm, the distribution of multiplicity extends down to the cutoff $n=0$, and small deviations from the Gaussian behavior start to appear. 
For these reasons, we exclude the most peripheral collisions in the following analysis. 
This is also motivated by the fact that fluctuations of $n$ for large $b$ are expected to be large, and large fluctuations are in general non-Gaussian.

A crucial quantity which we will use throughout this work is the position of the \textit{knee} of the distribution of $n$. 
We define it as the mean value of $n$ at zero impact parameter:\footnote{An alternative definition~\cite{ATLAS:2017zcm} is to define $n_{\rm knee}$ as the value of $n$ which minimizes the derivative $dP(n)/dn$, that is, the rightmost inflection point of $P(n)$ when plotted on a linear scale (as opposed to the logarithmic scale of Fig.~\ref{fig:allexp}). Both definitions are equivalent in the limit $\sigma(c_b)\to 0$.}
\begin{equation}
\label{defknee}
n_{\rm knee}\equiv \bar n(0).
\end{equation}
The observed distribution of $n$, $P(n)$, is eventually obtained by integrating Eq.~(\ref{nfixedb}) over $c_b$, i.e.,
\begin{equation}
\label{vzerodist}
P(n)=\int_0^1 P(n|c_b)dc_b. 
\end{equation}
In this paper, we determine smooth functions $\bar n(c_b)$ and $\sigma(c_b)$ such that $P(n)$ matches experimental data. 
This problem is underconstrained, in the sense that one cannot determine two unknown functions $\bar n(c_b)$ and $\sigma(c_b)$ from a single function $P(n)$. 
We shall argue that one can only constrain the mean $\bar n(c_b)$, and the value of the width for central collisions, $\sigma(0)$. 
Since the variation of $\sigma$ with $c_b$ cannot be determined from data alone, we test two different scenarios:
\begin{itemize}
\item{(A) $\sigma(c_b)=\sigma(0)\sqrt{\bar n(c_b)/\bar n(0)}$}
\item{(B) $\sigma(c_b)=\sigma(0)$.}
\end{itemize}
The first scenario, (A), assumes that the variance is proportional to the mean, which would be true if $n$ were the sum of contributions from independent nucleon-nucleon collisions. 
Scenario (B) is motivated by the observation that the width of the histograms observed in Fig.~\ref{fig:gaussian} varies little between $b=0$ and $b=8$~fm. 
We state that this is an artifact of the Monte Carlo model, where particle production is essentially determined by the participant nucleons. 
Since the number of participant nucleons is bounded by the total number of nucleons, fluctuations of $n$ are consequently reduced by the presence of this upper cutoff.
There is, however, no deep theoretical reason to believe that this particular feature of the Monte Carlo models is realistic. 

For each scenario, (A) or (B), we need a smooth function $\bar n(c_b)$, and a constant, $\sigma(0)$, such that $P(n)$ defined by Eqs.~(\ref{nfixedb}) and (\ref{vzerodist}) fits experimental data. 
We use the following functional form of $\bar n(c_b)$, which guarantees positivity:  
\begin{equation}
\label{fitfunction}
\bar n(c_b)=n_{\rm knee}\exp\left(-a_1 c_b-a_2 c_b^2-a_3 c_b^3\right).
\end{equation}
One could as well use other functional forms, requiring the fitting function to be a smooth, positive, monotonically decreasing function of $c_b$, with no singularities in the interval $0\le c_b<1$.
We carry out a five-parameter fit to $P(n)$ using Eqs.~(\ref{nfixedb}) and (\ref{vzerodist}), with parameters given by $n_{\rm knee}$, $a_1$, $a_2$, $a_3$ and $\sigma(0)$. 
To eliminate peripheral collisions from the fit, we only use values of $n$ above a cutoff  $n_{\min}$, which we specify in each case. 

\section{validation of the method}
\label{s:trento}

\begin{figure}[t!]
\begin{center}
\includegraphics[width=.9\linewidth]{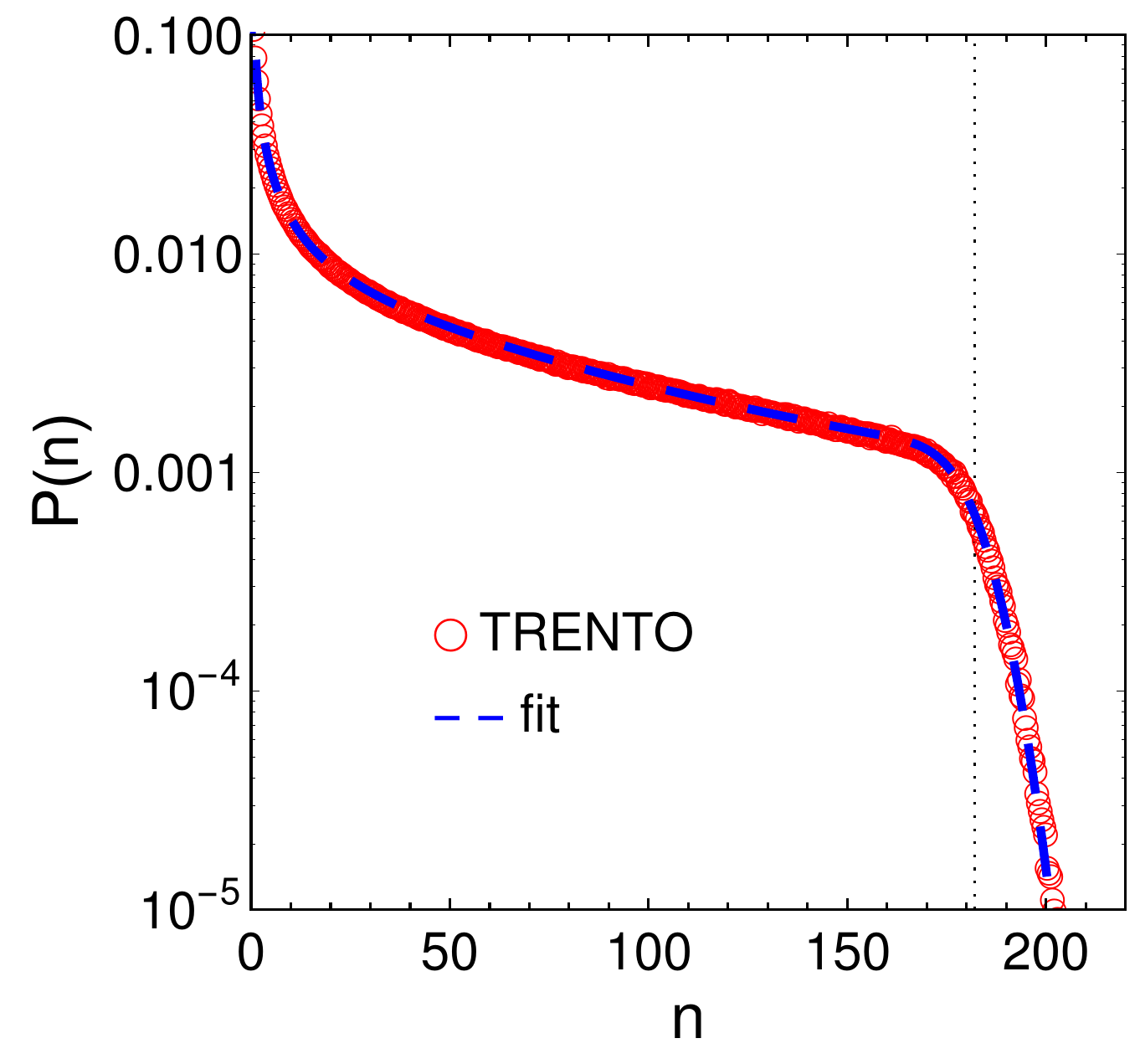}
\end{center}
\caption{(Color online) Circles show $P(n)$ obtained from the {\trento} model of initial conditions. The dashed line shows a fit of $P(n)$ using Eqs.~(\ref{nfixedb}) and (\ref{vzerodist}) in the scenario (A). The vertical line indicates the position of the knee [see Table~\ref{tabletrento}]. The quantity $n$ is in arbitrary units.
}
\label{fig:trentofit}
\end{figure}

We now validate this procedure of relating $n$ to $c_b$ using Monte Carlo simulations, where both $b$ and $n$ are known in each event.
We simulate Pb+Pb collisions at $~\sqrt{s}=2.76$~TeV using the same setup of \trento{} as in Fig.~\ref{fig:gaussian}, and again we use the entropy of each event to construct the probability distribution $P(n)$.
We generate $10^7$~events. 
We determine $c_b$ of each event by sorting events according to $b$, and $c$ by sorting them according to $n$. 
Symbols in Fig.~\ref{fig:trentofit} correspond to the distribution $P(n)$ from this Monte Carlo calculation. 

We now apply the fitting procedure described in the previous section to $P(n)$. 
For scenario (A), we vary the lower cutoff from $n_{\min}=1.8$ to $n_{\min}=9.1$, corresponding to centralities $c=80\%$ and $c=62\%$, and we find that the results of the fit are stable. 
The dashed line in Fig.~\ref{fig:trentofit} displays the fit corresponding to the larger value of $n_{\min}$.
The fit provides an excellent description of $P(n)$ in the model.
For case (B), fluctuations are larger and we use a larger cutoff, $n_{\min}=36$.
Above the cutoff, the two fits are indistinguishable in Fig.~\ref{fig:trentofit}.
It may seem surprising, at first sight, that two different parametrizations yield identical fits.
However, the only difference between (A) and (B) is the width of fluctuations away from central collisions. 
The effect of the fluctuations is to smear $P(n)$ around a central value, and this smearing has a small effect. 

The fit parameters are listed in Table~\ref{tabletrento}, for both scenarios.
The error on the knee (defined as the relative difference with the direct calculation, see below), $n_{\rm knee}$, returned by the fit is only $0.3\%$, while the error on the width $\sigma(0)$ is about 3\%. 
Interestingly, the two fitting procedures (A) and (B) return essentially the same values of these two quantities.\footnote{The larger difference in $a_2$ and $a_3$ is due to the different lower cutoff, $n_{\min}$, used for (B).}

\begin{figure}[t!]
\begin{center}
\includegraphics[width=.9\linewidth]{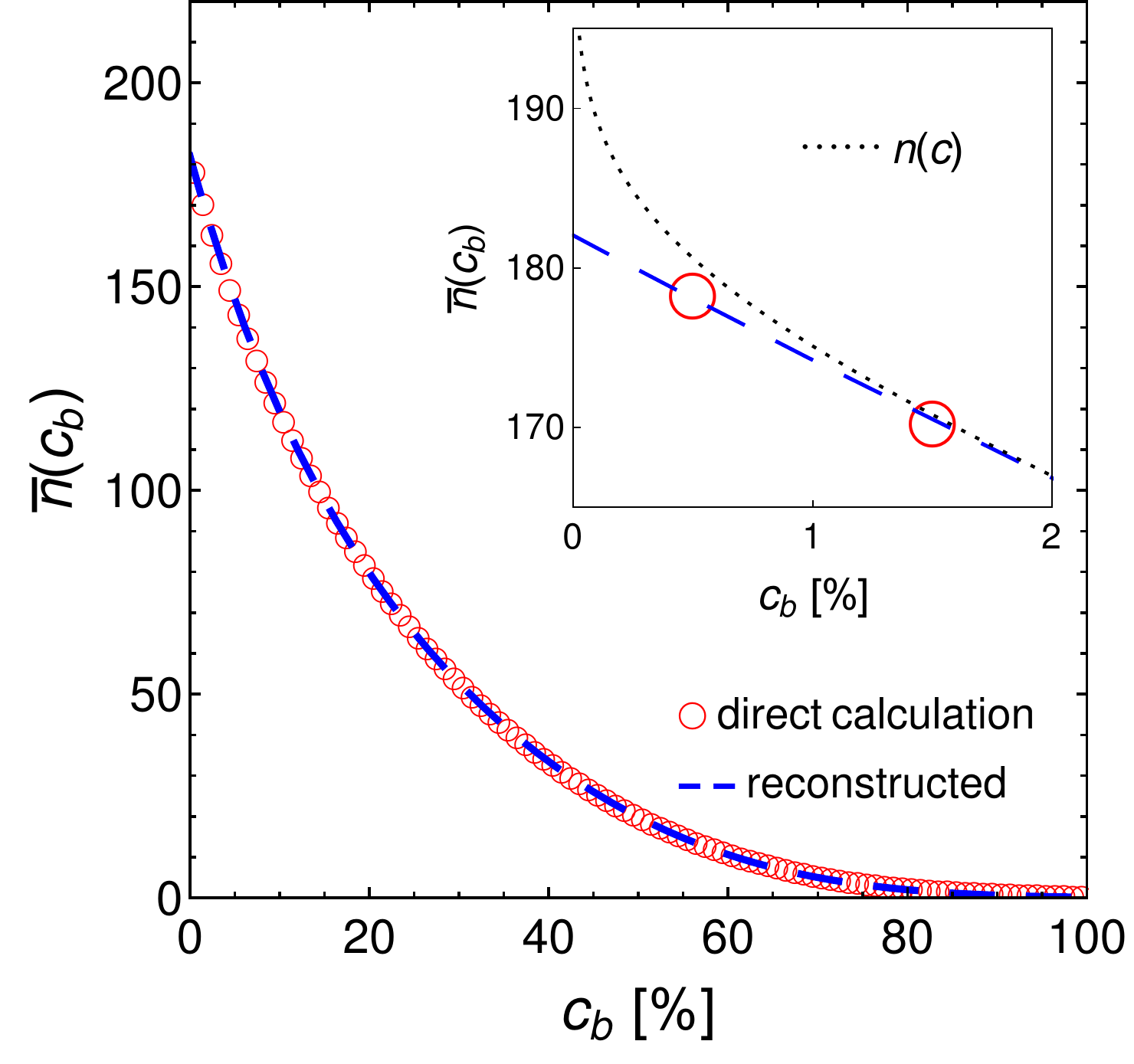}
\end{center}
\caption{(Color online) Mean value of $n$ versus $b$-centrality. 
Circles show results calculated directly by binning \trento{} results in $c_b$. The dashed line shows results reconstructed from the fit of $P(n)$ alone. The inset shows a zoom of the most central collisions, where we compare $\bar n (c_b)$ to $n(c)$ (dotted line).
}
\label{fig:trentoebar}
\end{figure}

In Fig.~\ref{fig:trentoebar}, we show $\bar n(c_b)$ returned by the fit, i.e., we calculate Eq.~(\ref{fitfunction}) using the parameters of Table~\ref{tabletrento} for scenario (A).
We display the comparison between this analytical estimate and the results directly obtained by binning \trento{} results in $c_b$, and computing the mean value of $n$ in each bin.
Agreement is within $0.5\%$, all the way up to 
$c_b=70\%$, which corresponds to a value of $n$ smaller than the lower cutoff applied to $P(n)$.
\begin{table}[b!]
\begin{tabular}{c|c|c|c}
\hline
\hline
&direct&fit (A)&fit (B)\cr
&&$\sigma(c_b)\propto\sqrt{\bar n(c_b)}$&$\sigma(c_b)=$const.\cr
\hline
$n_{\rm knee}$&$181.49\pm 0.13$ &$182.03$& $181.91$ \cr
$\sigma(0)$ &$7.79\pm 0.04$&$8.03$ & $8.09$\cr
$a_1$ &$4.36\pm 0.02$& $4.41$ &$4.48$\cr
$a_2$ &$-2.3\pm 0.1$& $-2.4$&$-3.0$\cr
$a_3$ &$4.8\pm 0.1$& $4.9$&$6.5$\cr
\hline
\hline
\end{tabular}
\caption{\label{tabletrento} 
Fit parameters obtained from fitting $P(n)$ in the {\trento} calculation (Fig.~\ref{fig:trentofit}). The direct calculation (first column) is the result obtained by binning the results in impact parameter (symbols in Fig.4), and fitting the resulting $\bar n(c_b)$ using Eq.(7) (see text). 
}
\end{table}
We also calculate $\sigma(0)$ directly, by generating $10^5$ collisions with $b=0$, and computing the standard deviation of $n$.

It is useful to note that approximate values of $\sigma(0)$ and $\bar n(c_b)$ can be read off directly from the shape of $P(n)$. 
Specifically, $\sigma(0)$ can be inferred from the width of the tail of the distribution on the right of the knee.  
$\bar n(c_b)$ is instead related to the shape of $P(n)$ left of the knee.
If $\sigma(c_b)$ is very small, the distribution of $n$ for fixed $b$ is very narrow, so that $c$ and $c_b$ tend to coincide. 
In this limit, $c(n)$ defined by Eq.~(\ref{defc}) is equal to the inverse function of $\bar n(c_b)$~\cite{Broniowski:2001ei}. 
$P(n)$ can thus be obtained by differentiating the centrality with respect to $n$, according to Eq.~(\ref{defc}): 
\begin{equation}
\label{blablabla}
P\left(n=\bar n(c_b)\right)\simeq -\left(\frac{d\bar n(c_b)}{dc_b}\right)^{-1}.
\end{equation}
In the inset of Fig.~\ref{fig:trentoebar}, we check the validity of this approximation in our \trento{} calculation by direct comparison of $\bar n (c_b)$ to $n(c)$.
We find that $\bar n (c_b)$ deviates from $n(c)$ only above $n\simeq 170$, corresponding to 1.5\% centrality.
This means that the effect of the finite width of the fluctuations on $P(n)$ is sizable only in the vicinity of the knee, i.e., in the most central collisions.

\begin{figure}[t!]
\begin{center}
\includegraphics[width=.793\linewidth]{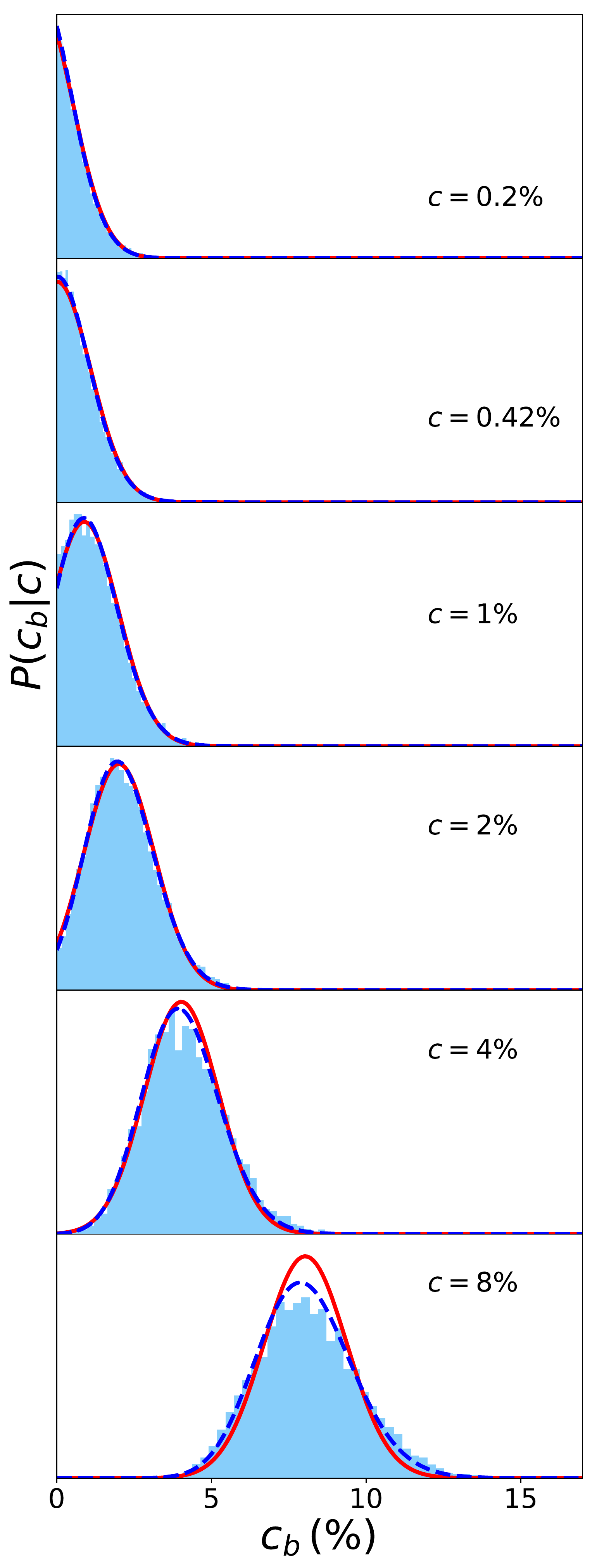}
\end{center}
\caption{(Color online)
Probability distribution of $b$-centrality in a very narrow centrality bin. Histograms correspond to a direct calculation using \trento{}.  Lines show the probability distributions of $c_b$ reconstructed through the Bayes' theorem, Eq.~(\ref{bayes}). Solid and dashed lines correspond to scenarios (A) and (B), respectively.  The second panel corresponds to $n= n_{\rm knee}$. Note that for $n \ge n_{\rm knee}$ (two uppermost panels), the most probable value of $c_b$ is $c_b=0\%$. Left of the knee, on the other hand, the most probable value of $c_b$ is $c$~\cite{Broniowski:2001ei}, as clearly visible in the two lowermost panels. All the curves displayed in this figure have area normalized to unity.
}
\label{fig:cbdist}
\end{figure} 
To complete our procedure, we show how one can reconstruct the distribution of impact parameter  for a given value of $n$, i.e., at a given value of centrality.
In the {\trento} model, this can done directly by sorting events into very narrow centrality bins, and then looking at the distribution of $c_b$ in each bin.
We show such histograms of $c_b$ for a few selected centrality bins in Fig.~\ref{fig:cbdist}.
Each panel corresponds to a bin of width $0.1\%$, centered around the displayed value of $c$ (for instance, the panel with $c=1\%$ shows the distribution of $c_b$ for $0.95\% < c < 1.05\%$.)

In an experimental situation where the impact parameter is not known, these distributions can be reconstructed from $P(n)$ using Bayes' theorem: 
\begin{eqnarray}
\label{bayes}
P(c_b|c)&=&P(c|c_b)\cr
&=&\frac{P(n|c_b)}{P(n)}, 
\end{eqnarray}
where, in the first line, we have used the property that the distribution of $c_b$ and $c$ are uniform, i.e., $P(c_b)=P(c)=1$. 
The distribution $P(n|c_b)$ is given by Eq.~(\ref{nfixedb}), where $\bar n(c_b)$ and $\sigma(c_b)$ can be obtained from the fitting procedure. 

One first needs to determine the value of $n$ corresponding to a given centrality, $c$. 
This can be done straightforwardly by using the fitting function, which offers a smooth interpolation of $P(n)$.\footnote{Experimental data typically present $P(n)$ in a bin center, and the centrality, $c$, at the boundary between two bins.}
Inserting Eq.~(\ref{nfixedb}) into Eq.~(\ref{vzerodist}), Eq.~(\ref{defc}) yields, after exchanging the order of integrals, 
\begin{equation}
\label{erfc}
c=\int_0^1 \frac{1}{2}~{\rm erfc}\left(\frac{n-\bar n(c_b)}{\sqrt{2}\sigma(c_b)}\right)dc_b ,
\end{equation}
where ${\rm erfc}(x)$ denotes the complementary error function.  
Eventually, once $n$ is determined, $P(c_b|c)$ is given by application of Bayes' theorem, Eq.~(\ref{bayes}), to $P(n|c_b)$ in Eq.~(\ref{nfixedb}).
The reconstructed distributions are shown as lines in Fig.~\ref{fig:cbdist}. 
Scenarios (A) and (B) are represented by solid and dashed lines, respectively.
Both are in perfect agreement with the direct calculation up to $4\%$ centrality. 
Discrepancies between both scenarios and the direct calculation start to appear around $c=8\%$, meaning that our approximated formulas for $\sigma(c_b)$ starts to break down around $c\sim10\%$ in this model calculation.
In a sense, this is a consequence of the fact that the variation of $\sigma$ with $c_b$ can not be inferred from the sole $P(n)$.


\begin{table}[t!]
\begin{tabular}{c|c|c|c|c}
\hline
\hline
&STAR~\cite{Adler:2001yq}&ALICE~\cite{Abelev:2013qoq}&ATLAS~\cite{ATLAS:2011ah}&CMS~\cite{Chatrchyan:2012vq}\cr
\hline
$n_{\rm knee}$&296.8 &20406&3.575 & 119.03 \cr
$\sigma(0)$ &21.5 &731 &0.113 & 3.82 \cr
$a_1$ &3.55&4.11&4.05 & 4.09 \cr
$a_2$ &0.8& -1.9&-1.5 & -1.8 \cr
$a_3$ &1.6& 4.4&4.1 & 4.2\cr
\hline
\hline
\end{tabular}
\caption{\label{tableexp} 
Values of fit parameters for several experiments. 
}
\end{table}

\section{Application to data}
\label{s:data}

Our method of relating $c$ to $c_b$ being validated on model calculations, we apply now the fitting procedure to experimental data.
We fit the experimental curves of $P(n)$ shown in Fig.~\ref{fig:allexp} using Eq.~(\ref{vzerodist}).
The results shown in this section correspond to scenario (A), with a lower centrality cutoff.
We have checked that results are stable if one varies the cutoff, or if one uses scenario (B). 
The values of the fit parameters extracted from experimental data are given in Table~\ref{tableexp}, for all the analyzed experiments. 
The values of $n_{\rm knee}$ and $\sigma(0)$ are in the same units as $n$, which are usually arbitrary, and vary from one experiment to the other.
The other fit parameters are dimensionless. 
\begin{figure}[t!]
\begin{center}
\includegraphics[width=.8\linewidth]{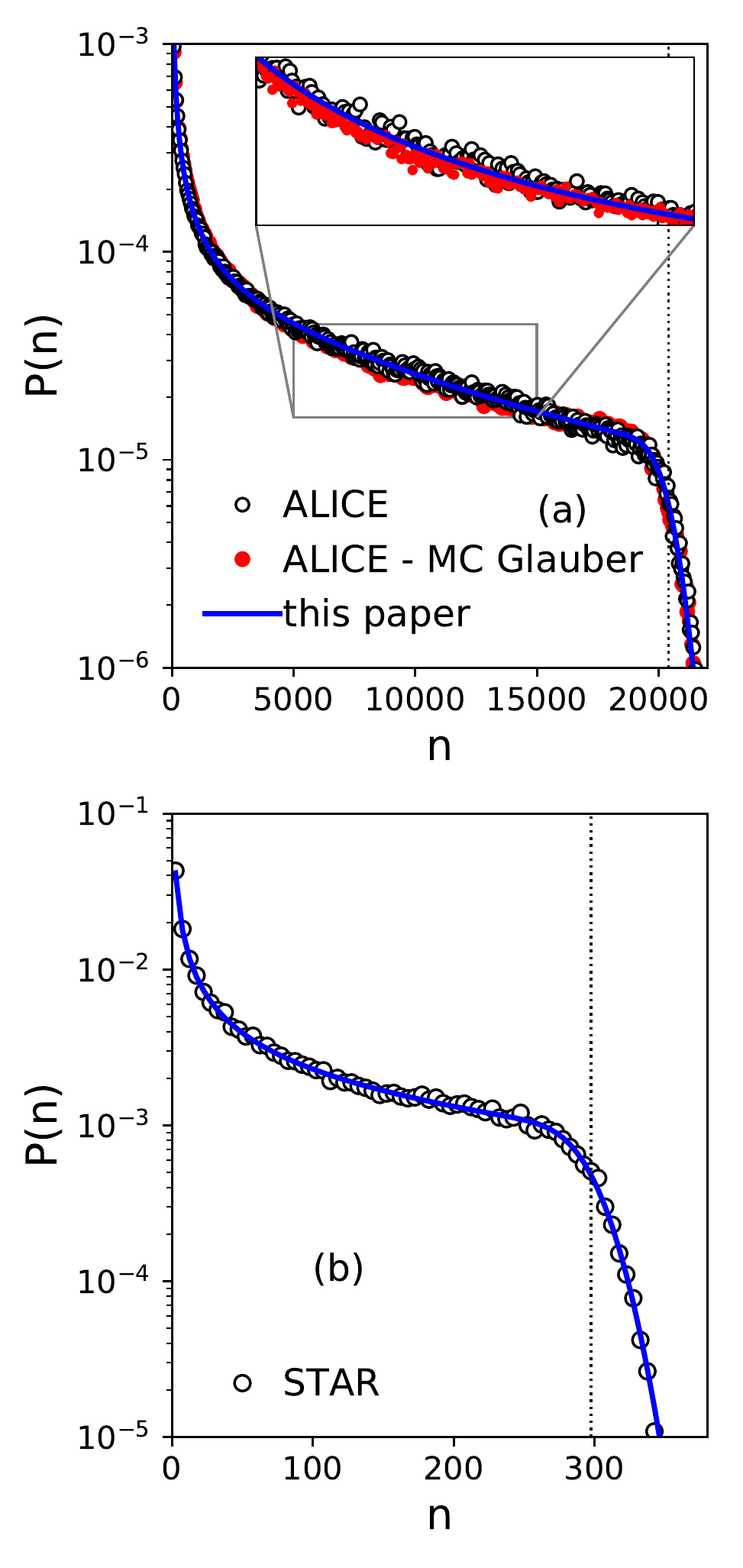}
\end{center}
\caption{(Color online) (a) Empty symbols: distribution of the VZERO amplitude (denoted by $n$), used by the ALICE collaboration to determine the collision centrality~\cite{Abelev:2013qoq}.      
Full symbols: Glauber model used by the ALICE collaboration to fit the measured distribution. 
Line: Fit of data provided by Eq.~(\ref{vzerodist}). 
The inset is a zoom of the central part of the histogram. The quantity $n$ is in arbitrary units.
 (b) Symbols: STAR data \cite{Adler:2001yq}. Line: fit using Eq.~(\ref{vzerodist}). 
The vertical line in both panels indicates the position of the knee.
}
\label{fig:alicefit}
\end{figure}

In Fig.~\ref{fig:alicefit}--(a) we display raw data on $P(n)$ (empty symbols) measured by the ALICE Collaboration, together with our fit (line).
The $\chi^2$ per degree of freedom of the fit is $1.2$.
Along with data, we plot also the distribution of $n$ provided by the Monte Carlo Glauber model used by the ALICE Collaboration (full symbols) to fit their $P(n)$, and, consequently, to perform the sorting of events into centrality bins.
A remarkable outcome of our fitting method is that it provides a description of experimental data which is better than that provided by the Glauber model tuned to ALICE data, as evident from the inset of Fig.~\ref{fig:alicefit}--(a), where we zoom in the central body of the histogram.
Panel (b) displays the fit to STAR data, which is as good as the fit to ALICE data. 
The fits of ATLAS and CMS data (not shown) are of the same quality. 

\begin{table}[b!]
\begin{tabular}{c|c}
\hline
\hline
Experiment& $c_{\rm knee}$\cr
\hline
STAR~\cite{Adler:2001yq}& $0.81\% \pm 0.10\%$ \cr
ALICE~\cite{Abelev:2013qoq}&$0.349\% \pm 0.023\%$ \cr
ATLAS~\cite{ATLAS:2011ah}&$0.313\%\pm 0.011\%$\cr
CMS~\cite{Chatrchyan:2012vq}&$0.314\%\pm 0.040\%$\cr
\hline
\hline
\end{tabular}
\caption{\label{cknee} 
Fraction of events above the knee for various heavy-ion experiments. 
}
\end{table}

A convenient measure of the accuracy of the centrality determination is the fraction of events above the knee of the distribution, that is, the centrality of the knee, which we denote by $c_{\rm knee}$.
It is obtained by replacing $n$ with $n_{\rm knee}$ in Eq.~(\ref{erfc}).
Its value for each of the considered experiments is given in Table~\ref{cknee}.
The recovered ordering of $c_{\rm knee}$ among the different collaborations is consistent with the curves shown in Fig.~\ref{fig:allexp}, because $c_{\rm knee}$ is proportional to the width of the tail of $P(n)$.
We now derive a approximate expression of $c_{\rm knee}$ which provides a simple way to relate it to the fit parameters of Table~\ref{tableexp}. 

If $n=n_{\rm knee}$, only small values of $c_b$ contribute to the integral in Eq.(\ref{vzerodist}).  
Therefore, we can expand $n(c_b)$ to first order in $c_b$ in Eq.~(\ref{nfixedb}), obtaining $n(c_b)-n_{\rm knee}\simeq c_b (d\bar n/dc_b)|_{c_b=0}$. 
Neglecting the variation of the width with $c_b$, i.e., $\sigma(c_b)\approx\sigma(0)$, and replacing $\bar n (c_b)$ in Eq.~(\ref{erfc}), one obtains
\begin{equation}
c_{\rm knee}=-\frac{\sigma(0)}{\left.\frac{d\bar n}{dc_b}\right|_{c_b=0}\sqrt{2\pi}}=
\frac{\sigma(0)}{n_{\rm knee} a_1\sqrt{2\pi}},
\end{equation}
where, in the last equality, we have used Eq.~(\ref{fitfunction}).
We have checked that, using this simple estimate, the values of $c_{\rm knee}$ shown in Table~\ref{tableexp} are reproduced to a good accuracy, within 1\%.
\begin{figure}[t!]
\begin{center}
\includegraphics[width=.8\linewidth]{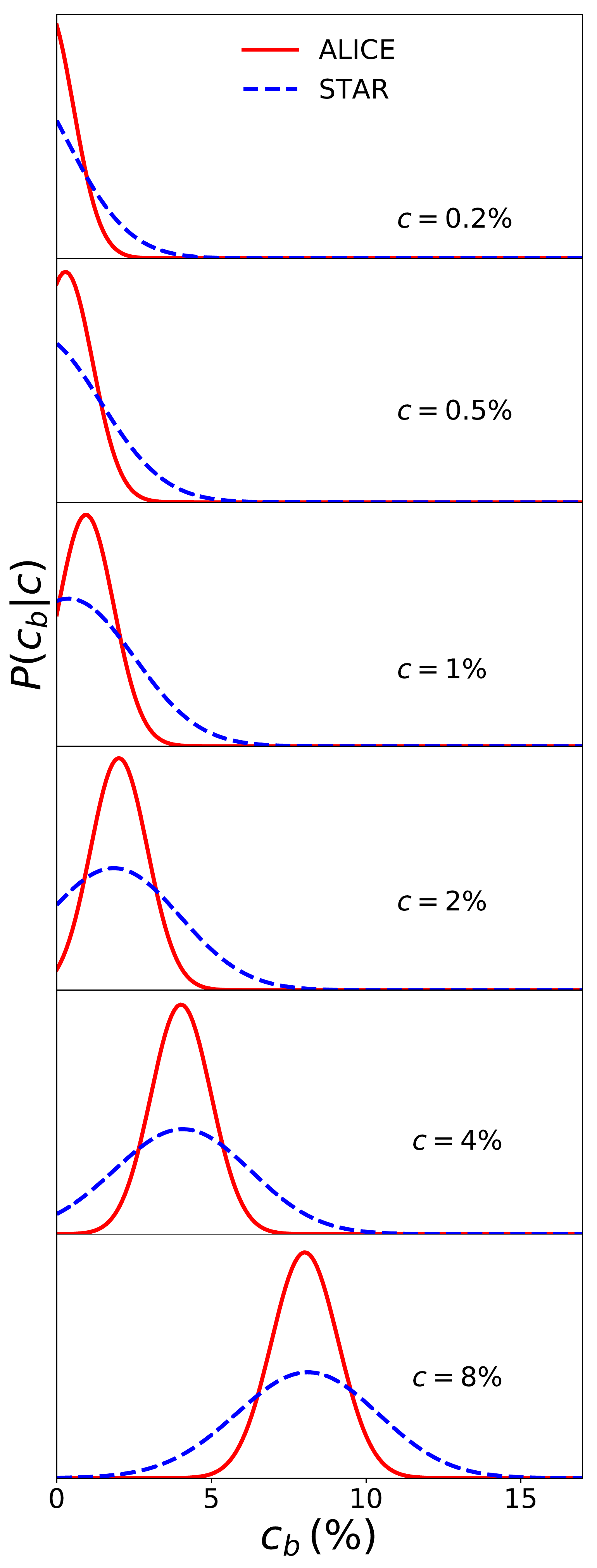}
\end{center}
\caption{(Color online)
Probability distributions of $b$-centrality for selected values of the
centrality reconstructed from STAR (dashed lines) and ALICE (solid lines) data using Bayes' theorem. All the curves shown in this figure have area normalized to unity.
}
\label{fig:cbdist2}
\end{figure}

\begin{figure}[t!]
\begin{center}
\includegraphics[width=\linewidth]{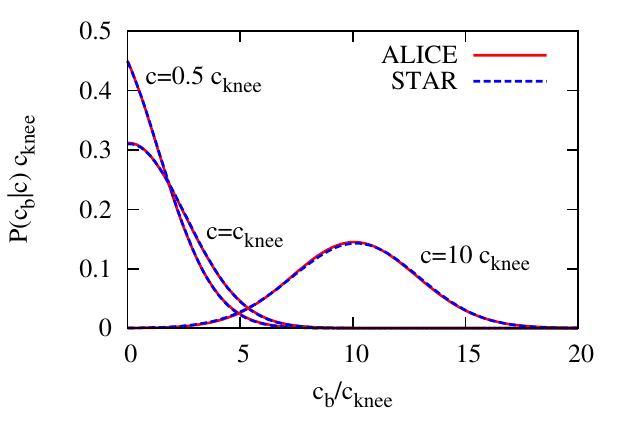}
\end{center}
\caption{(Color online)
Probability distribution of $c_b/c_{\rm knee}$ extracted from STAR (solid lines) and ALICE (dashed lines) data for three selected
values of $c/c_{\rm knee}$.  
\label{fig:cbknee}
}
\end{figure}

Ultimately, using Bayes' theorem, we can use the fits of experimental data to reconstruct the distribution of impact parameter for a fixed centrality. 
Figure~\ref{fig:cbdist2} presents the distribution of $c_b$ for a few selected values of centrality percentile. 
The distributions obtained by fitting data of the LHC Collaborations are very similar, therefore, we present only a comparison between the curves derived from ALICE and STAR data.
We find that the distribution of $c_b$ extracted from STAR data is much broader than the one extracted from ALICE data. 
This is a direct consequence of the wider tail of $P(n)$ measured by the STAR Collaboration [see Fig.~\ref{fig:allexp}], which in turn can be ascribed to the smaller multiplicity seen by the detector, as explained in Sec.~\ref{s:defcent}. 
Note that, as expected, the distributions extracted from STAR and ALICE data are almost identical if one rescales both $c$ and $c_b$ by $c_{\rm knee}$, as we show in Fig.~\ref{fig:cbknee}.

\section{Conclusion}
We have shown that even though the impact parameter of a nucleus-nucleus collision is not a measurable quantity, precise information about impact parameter is contained in available experimental data. 
We have delineated a procedure allowing to reconstruct accurately the probability distribution of impact parameter at a given centrality (as defined experimentally by a multiplicity or a transverse energy), up to $5-10\%$ centrality.  
This reconstruction does not involve the concept of participant nucleons, or any microscopic model of the collision.
Moreover, it is independent of the detector efficiency and other detector-related effects (e.g., resolution, saturation due to multiple hits), as long as the calibration of the centrality is correct and the detector response is stable throughout the run.
Its sole inputs are the distribution $P(n)$, where $n$ is the observable used to determine the collision centrality, along with the assumption that the distribution of $n$ for a fixed $b$ is Gaussian.
We stress that this assumption is solidly rooted in the central limit theorem.

The fraction of events above the knee of $P(n)$, $c_{\rm knee}$, provides a simple measure of the precision of the centrality determination.
It is below $0.4\%$ at LHC, and twice larger at STAR. 
We have shown that impact parameter fluctuations in the $0-10\%$ most central collisions are essentially determined by this quantity alone.

We have shown that the mean value of $n$ at a fixed impact parameter can be reconstructed accurately up to $70\%$ $b$-centrality.
The standard deviation of $n$ around the mean can instead be reconstructed only for $b=0$: Its centrality dependence cannot be inferred from $P(n)$ alone.  
The mean and the standard deviation of $n$ at fixed $b$ are more natural quantities from a theory point of view than $P(n)$, because they can be directly obtained in a model by fixing the impact parameter.  
While the standard deviation may depend on detector details (in particular, purely statistical fluctuations are larger in relative value if the acceptance is smaller), the mean, $\bar n (c_b)$, provides a robust quantity for model comparisons. 

It would be useful to extend this study to proton-nucleus collisions. 
However, our assumption that $n$ has Gaussian fluctuations is not satisfied in model calculations, even for central collisions. 
We have checked that the fit procedure is less successful in 
describing $p$+Pb collision data at $~\sqrt[]{s}=5.02~{\rm TeV}$~\cite{Adam:2014qja}. 
In particular, the tail of the distribution of $P(n)$ is not as well reproduced, because it is exponential rather than  Gaussian. 
The origin of this exponential tail is well understood theoretically~\cite{Liou:2016mfr}. 
This difference must be taken into account in order to extend our study to smaller collision systems. 

\section*{Acknowledgements}
We thank Alberica Toia for providing us with ALICE data, and Jiangyong Jia for providing us with ATLAS data. 
We thank Jean-Paul Blaizot, Matt Luzum, Sasha Milov, St\'ephane Munier, Art Poskanzer, and Lenka Zdeborova for useful discussions.

\end{document}